# Role of Fe substitution on the anomalous magnetocaloric and magnetoresistance behavior in Tb(Ni$_{1-x}$Fe$_x$)$_2$ compounds

Niraj K. Singh[1], K. G. Suresh[1*], D. S. Rana[2], A. K. Nigam[2] and S. K. Malik[2]

[1]Indian Institute of Technology Bombay, Mumbai-400076, India

[2]Tata Institute of Fundamental Research, Homi Bhabha Road, Mumbai-400005, India

Corresponding author (email: suresh@phy.iitb.ac.in)

**Abstract.** We report the magnetic, magnetocaloric and magnetoresistance results obtained in Tb(Ni$_{1-x}$Fe$_x$)$_2$ compounds with x=0, 0.025 and 0.05. Fe substitution leads to an increase in the ordering temperature from 36 K for x=0 to 124 K for x=0.05. Contrary to a single sharp MCE peak seen in TbNi$_2$, the MCE peaks of the Fe substituted compounds are quite broad. We attribute the anomalous MCE behavior to the randomization of the Tb moments brought about by the Fe substitution. Magnetic and magnetoresistance results seem to corroborate this proposition. The present study also shows that the anomalous magnetocaloric and magnetoresistance behavior seen in the present compounds is similar to that of Ho(Ni,Fe)$_2$ compounds.

## 1. INTRODUCTION

The property of the magnetic materials to heat up or cool down when they are subjected to a varying magnetic field under adiabatic conditions is known as magnetocaloric effect (MCE). Generally, the MCE is measured as isothermal magnetic entropy change and/or as adiabatic temperature change. Materials with considerable MCE over a wide temperature span find applications as active materials in magnetic refrigerators. The discovery of giant MCE in a few intermetallic compounds [1-3] has led to an intense research towards the development of a magnetic refrigerator for room temperature applications. However, apart from its potential in the near room temperature region, the magnetic refrigeration technology could play a vital role in cryogenic applications spread over a wide temperature range [4, 5]. For example, the principle of magnetic refrigeration is being used in devices like gas like liquefiers, cryo-coolers etc [6] Furthermore, some recent reports suggest that magneto-thermal properties associated with magnetic materials could also be exploited in the fabrication of heat pumps. Therefore, the search for novel magnetic materials and studies on their MCE behavior are of great relevance today.

Manifestation of diverse magnetic properties by the rare earth (R) transition metal (TM) intermetallic compounds in general and the occurrence of giant MCE in materials such as $Gd_5(Si,Ge)_4$ in particular have made R-TM compounds as the natural probe for the fundamental studies as well as for applications based on MCE [1, 7-9]. Among the various R-TM intermetallics, $RCo_2$ (R=Er, Ho and Dy) compounds are known to exhibit considerable MCE due to the first order transition at their ordering temperature ($T_C$) [8]. Another series which is quite promising is RNiAl. Large table-like MCE observed in many compounds of this series is attributed to the presence of multiple magnetic transitions in this series of compounds [9]. Systems based on $La(Fe,Si)_{13}$ also exhibit significant MCE, enabling them to be considered as potential refrigerants [10]. Most of the materials studied so far show considerable MCE close to their magnetic ordering temperatures, resulting in a single peak in the temperature variation of MCE plots. However, there exist a few materials, typical examples being $GdMn_2$ and $Gd_3Al_2$, which show a double-peak MCE behavior [11, 12]. Such materials are of interest in the field of magnetic refrigeration since they may lead to considerable MCE over a wide range of temperature.

As part of the program to develop novel magnetocaloric materials suitable for different temperature ranges and to study the correlation between the magnetism and MCE, we have been focusing our studies on various $RTM_2$ Laves phase compounds [13-15]. Recently, we found that Fe substitution results in an anomalous magnetocaloric effect and magnetoresistance (MR) in $Ho(Ni,Fe)_2$ system [15]. In the case of MCE, the anomaly was manifested in the form of a broad double-peak structure in the Fe-substituted compounds. This behavior was attributed to the randomization of the moments in the Ho sublattice, brought about by the Fe substitution, at low temperatures. Magnetization and MR results were found to corroborate the presumption of randomization. In order to ascertain whether this behavior is linked to a particular rare earth, we have taken up another system namely $Tb(Ni_{1-x}Fe_x)_2$ [with x= 0, 0.025 and 0.05] and studied MCE and MR, the results of which are presented in this paper.

## 2. EXPERIMENTAL DETAILS

Methods of preparation of the compounds Tb(Ni$_{1-x}$Fe$_x$)$_2$, with x= 0, 0.025 and 0.05 and their characterization have been reported elsewhere [15]. Magnetization measurements in the temperature range of 2-240 K and up to a maximum field of 50 kOe, were carried out using a vibrating sample magnetometer (VSM, Oxford instruments). The ac magnetic susceptibility was measured using a Physical Property Measurement System (PPMS, Quantum Design) in the temperature range of 2-200 K. Heat capacity measurements, in the temperature range 2- 280 K and in fields up to 50 kOe, were performed using the relaxation method in a PPMS. The magnetoresistance has been calculated by measuring the electrical resistivity in fields up to 50 kOe in the temperature range 2-300 K, using the linear four-probe technique.

## 3. RESULTS AND DISCUSSION

The Rietveld refinement of the room temperature powder x-ray diffractograms confirms that all the compounds crystallize in MgCu$_2$-type Laves phase cubic structure (space group *Fd3m*, number 227). The refined lattice parameters (*a*) are 7.154 ±0.001, 7.183 ± 0.001 and 7.201 ± 0.001 Å, for the compounds with x= 0, 0.025 and 0.05, respectively.  The temperature (T) dependence of magnetization (M) data, collected in an applied field (H) of 500 Oe, reveals that the ordering temperature of the compounds with x=0, 0.025 and 0.05 are 36 K, 87 K and 124 K, respectively. Figure 1 shows the M-T plot of Tb(Ni$_{0.975}$Fe$_{0.025}$)$_2$ obtained in a field of 500 Oe, under zero field cooled (ZFC) and field cooled (FC) conditions. A similar plot was obtained for Tb(Ni$_{0.95}$Fe$_{0.05}$)$_2$ as well. Large thermo-magnetic irreversibility between ZFC and FC magnetization data seen in these compounds is attributed to the domain wall pinning effect [13]. Another feature worth noting in this figure is that the FC magnetization of Tb(Ni$_{0.975}$Fe$_{0.025}$)$_2$, at low temperatures, shows a rather faster reduction with increase in temperature. This behavior is found to be even more prominent in the case of Tb(Ni$_{0.95}$Fe$_{0.05}$)$_2$. The inset of figure1 shows the ZFC-FC magnetization plots of TbNi$_2$. As can be seen from the inset, TbNi$_2$ shows an anomaly in the form of a kink at about 15 K, both in the FC and ZFC magnetization.

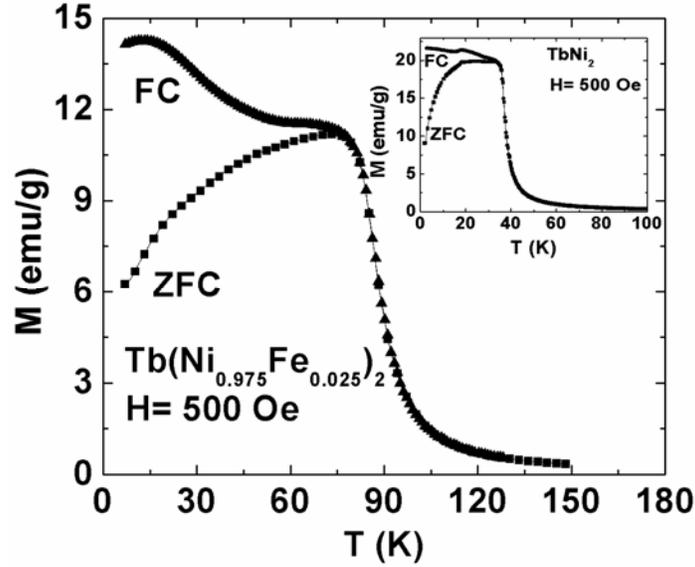

Figure 1 Temperature variation of ZFC and FC magnetization data of Tb(Ni$_{0.975}$Fe$_{0.025}$)$_2$, obtained in an applied field of 500 Oe. The inset shows the M-T plot of TbNi$_2$, realized under similar conditions.

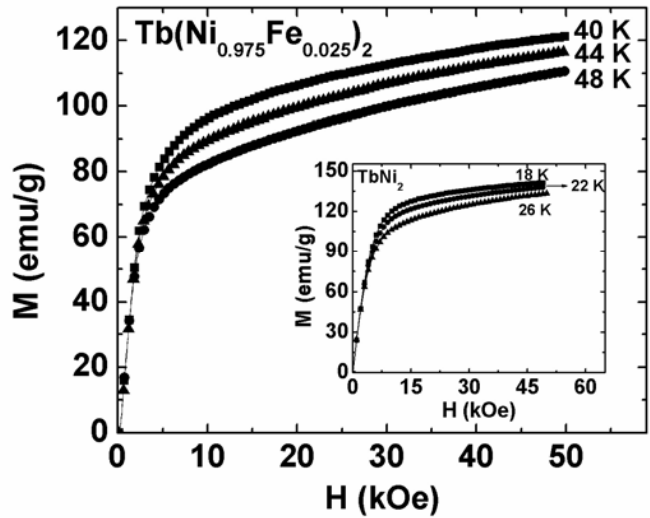

Figure 2 M-H isotherms of Tb(Ni$_{0.975}$Fe$_{0.025}$)$_2$, obtained at about half of the ordering temperature. The inset shows the corresponding plots of TbNi$_2$.

We have also studied the field dependence of magnetization data at various temperatures. Of particular interest is the M-H isotherms obtained at temperatures close to about half of the ordering temperature for all the compounds and the representative plots for Tb(Ni$_{0.975}$Fe$_{0.025}$)$_2$ are given in figure. 2. The M-H isotherms of $TbNi_2$ (at temperatures close to $T_C/2$) are given as an inset of figure. 2. It may be noticed that the M-H isotherms of $TbNi_2$, at high fields show a better saturation tendency compared to that of $Tb(Ni_{0.975}Fe_{0.025})_2$. It should be kept mind that such a non-saturation could arise from the change in crystalline electric field (CEF) effects brought about by Fe substitution. However, the fact that the M-H isotherm of $Tb(Ni_{0.975}Fe_{0.025})_2$ at 4 K saturates in the same way as $TbNi_2$ may suggest that the CEF is not altered considerably by Fe substitution. Therefore, the above observations may be indicative of the presence of a larger magnetic disorder in the Fe substituted compounds as compared to that of $TbNi_2$, in the same temperature range of $T/T_C$.

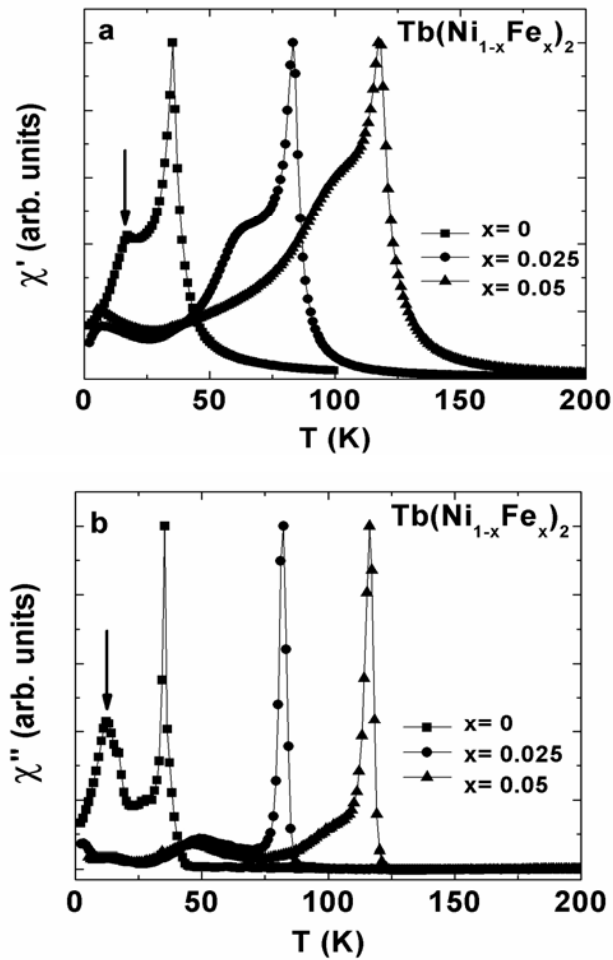

Figure 3 Temperature dependence of real (a) and imaginary (b) parts of normalized ac susceptibility of $Tb(Ni_{1-x}Fe_x)_2$ compounds obtained under zero dc bias field. The frequency of measurement is 333 Hz. The arrows indicate the spin reorientation transitions seen in $TbNi_2$.

To further probe the magnetic state of these compounds, we have carried out the ac magnetic susceptibility ($\chi_{ac}$) measurements with and without a dc bias field, in all these compounds. Figures 3a and 3b show the temperature variation of the real ($\chi'_{ac}$) and the imaginary ($\chi''_{ac}$) parts of ac susceptibility normalized to the respective maximum values. It is clear from Figure 3a that in all the compounds $\chi'_{ac}$ shows a peak at temperatures close to $T_C$. In the case of TbNi$_2$, apart from the peak at $T_C$, another peak is observed at about 15 K, which is consistent with the observation seen in the M-T data. We attribute this peak to the spin reorientation transition reported in this compound [16]. Another interesting observation from these plots is the broadening of the peaks in the Fe substituted compounds. It is to be noted that though the FC magnetization data of the Fe substituted compounds show some unusual reduction in the magnetization at low temperatures, no visible transition has been observed in the $\chi'_{ac}$-T plot. However, the $\chi''_{ac}$-T plot of these compounds show a secondary peak at about 50 K.

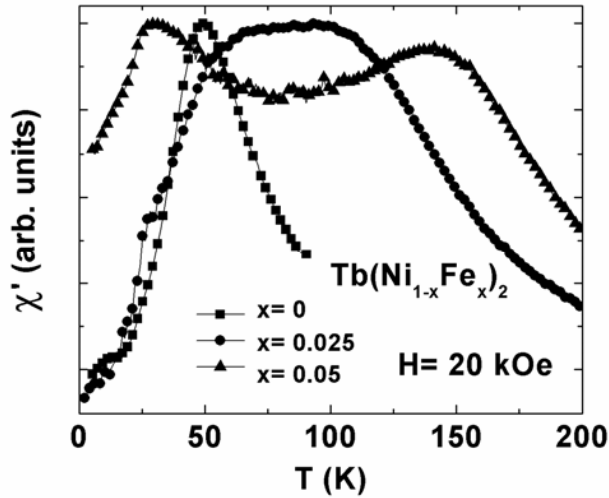

Figure 4 Temperature dependence of the real part of normalized ac susceptibility of Tb(Ni$_{1-x}$Fe$_x$)$_2$ compounds obtained under a dc bias field of 20 kOe. The frequency of measurement is 333 Hz.

In order to understand the nature of the low temperature magnetic state, we have obtained the $\chi'_{ac}$-T plot under a dc bias field of 20 kOe as well (Figure 4). It can be seen that the peak corresponding to the spin reorientation transition (~15 K) for TbNi$_2$ is absent in the presence of the dc bias field. It has indeed been reported that a field of 10 kOe completely suppresses the spin reorientation transition in TbNi$_2$ [16]. Furthermore, one can see that the peak corresponding to the ordering temperature has shifted from 36 K to about 50 K, as the dc bias is applied. Though the sharpness of the peak in TbNi$_2$ is almost unchanged with the dc bias, the width of the peaks has increased quite considerably in the case of the Fe-substituted compounds.

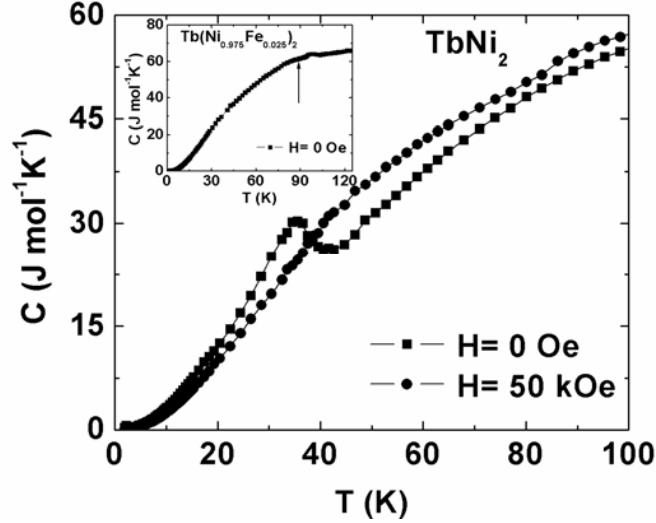

Figure 5 Heat capacity of TbNi$_2$ as a function of temperature in fields H=0 and 50 kOe. The inset shows the variation of zero-field heat capacity in Tb(Ni$_{0.975}$Fe$_{0.025}$)$_2$. The arrow in the inset shows the ordering temperature determined from the magnetization data.

Figure 5 shows the temperature variation of heat capacity (*C*) for TbNi$_2$ in fields *H*=0 and 50 kOe. The zero-field heat capacity data shows a peak, which nearly coincides with the $T_C$ observed from the M-T data. It may be noticed from the figure that, in the case of TbNi$_2$, the peak associated with the magnetic transition in the C-T plot starts at temperatures well above $T_C$. This may be attributed to the spin fluctuations, arising from the presence of magnetic polarons, above $T_C$. Such a behavior has been observed in other intermetallic compounds also [17, 18]. However, in the iron substituted compounds no peak in heat capacity could be seen even in zero field (inset in Figure 5). Since the ordering temperatures of the iron-substituted compounds are considerably higher than that of TbNi$_2$, the relatively larger lattice and electronic contributions to the total heat capacity presumably mask any weak peak due to the magnetic contribution. A similar behavior has been observed in Ho(Ni,Fe)$_2$ system [15].

The magnetocaloric properties of Tb(Ni$_{1-x}$Fe$_x$)$_2$ compounds, both from the magnetization data as well as from heat capacity data, have been determined using the methods reported earlier [15, 19]. Figure 6a shows the temperature variation of isothermal magnetic entropy change ($\Delta S_M$) using the C-T data obtained in H= 0 and 50 kOe, for all the compounds. $\Delta S_M$ values calculated from the C-H-T data were found to be in close agreement with those obtained from the M-H-T data. It can be seen from the figure that $\Delta S_M$ *vs*. T plot for TbNi$_2$, calculated for a field change ($\Delta H$) of 50 kOe, shows a maximum near T$_C$ with a value 4 Jmol$^{-1}$K$^{-1}$. The experimentally observed $\Delta S_M$ for TbNi$_2$ is less than the theoretically predicted [20] value of ~7 Jmol$^{-1}$K$^{-1}$. The temperature variation of $\Delta S_M$ for Tb(Ni$_{0.975}$Fe$_{0.025}$)$_2$ show two broad peaks. Apart from a weak primary peak near T$_C$, a secondary peak at temperatures much below the T$_C$ is also observed. In the case of Tb(Ni$_{0.95}$Fe$_{0.05}$)$_2$, a single broad peak with some signature of a small kink (close to T$_C$) is observed. The maximum value $\Delta S_M$ ($\Delta S_M^{max}$), for Tb(Ni$_{0.975}$Fe$_{0.025}$)$_2$, is found to be ~2.2 Jmol$^{-1}$K$^{-1}$ at about 43 K whereas for Tb(Ni$_{0.95}$Fe$_{0.05}$)$_2$, it is ~1.4 Jmol$^{-1}$K$^{-1}$ at about 53 K. The double-peak MCE behavior seen here is quite similar to that reported in Ho(Ni,Fe)$_2$ system [15]. Figure 6b shows the temperature variation of the adiabatic temperature change ($\Delta T_{ad}$), calculated from the C-H-T data. Comparing figures 6a and 6b, one can see that the temperature variation of $\Delta T_{ad}$ is identical to that of $\Delta S_M$, except for a well-defined peak below 5 K, observed in all the three compounds. Such a behavior in the temperature variation of $\Delta T_{ad}$ has been reported for other intermetallic compounds and may be attributed to the crystal field effect [21, 22].

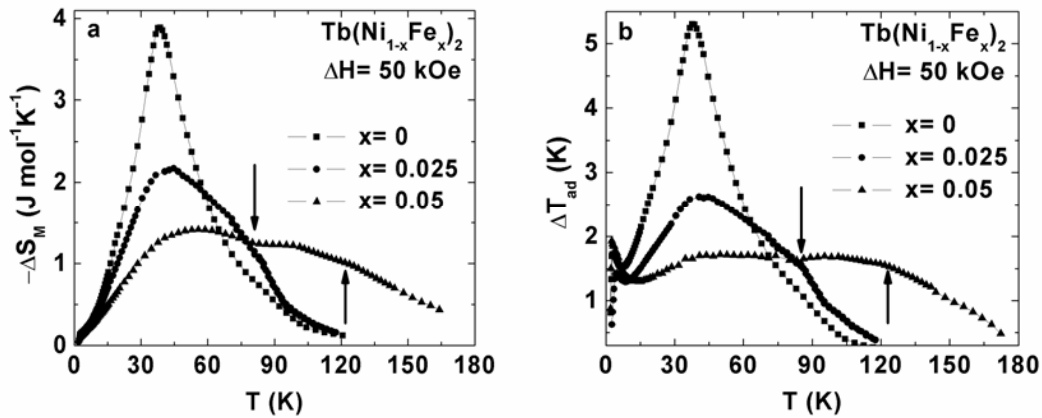

Figure 6. The temperature variation of magnetocaloric effect of Tb(Ni$_{1-x}$Fe$_x$)$_2$ compounds, in terms of isothermal magnetic entropy change (a) and adiabatic temperature change (b), for a field change of 50 kOe. The arrows in the figure indicate the ordering temperatures.

The presence broad peak in $\Delta S_M$ *vs*. T plot of the iron-substituted compounds indicates the existence of some degree of magnetic randomness, at low temperatures, in these compounds. It may be recalled here that the M-H isotherms of the Fe-substituted compounds, obtained at about half of the ordering temperature, indeed show some signature of a possible magnetic randomness at low temperatures (see Figure 2). Relatively broad peaks seen in the $\chi_{ac}$-T plots also indicate the possibility of randomness at low temperatures, in the case of Fe-substituted compounds. It is well known that the Ni sublattice in RNi$_2$ compounds is almost nonmagnetic [23] and hence the randomness cannot be attributed to Ni moments. Taking the maximum *J* value of 5/2, as per Hund's rule, the theoretical magnetic entropy [*Rln(2J+1)*] associated with the Fe sublattice in the compounds with x=0.025 and 0.05 are found to be 0.75 and 1.49 Jmol$^{-1}$K$^{-1}$ respectively. By comparing the theoretical values with the observed low temperature MCE peak values, one can rule out the Fe contribution towards the low temperature MCE peak. Therefore, the randomness occurring at low temperatures must be solely associated with the Tb sublattice. From a similar comparison of $\Delta S_M$ values associated with the primary peak with that of the theoretical magnetic entropy of the Fe sublattice, and taking into account the fact that in any magnetocaloric process only a fraction of the magnetic entropy is utilized [24], it can be inferred that $\Delta S_M$ values associated with the $\Delta S_M$ peak near T$_C$ could not arise from the Fe sublattice alone. Therefore, it appears that the randomness starts at low temperatures and continues upto T$_C$. This may be the reason for considerable MCE in the region between the two peaks. We have earlier attributed a similar MCE behavior observed in Fe substituted HoNi$_2$ compounds to a possible randomization of Ho moments [15]. The randomization was thought to be a result of the change in the local anisotropy brought about by the Fe substitution.

The MCE behavior of Tb(Ni$_{1-x}$Fe$_x$)$_2$ has also been studied for a field change of 10 kOe. The temperature variation of MCE of TbNi$_2$, calculated for $\Delta H$= 10 kOe is found to be similar to that obtained for 50 kOe, except for the peak values. However, in sharp contrast, in the Fe-substituted compounds the temperature variation of MCE obtained for $\Delta H$= 10 kOe does not show the secondary peak. This indicates that a critical field (more than 10 kOe) is needed to bring about considerable ordering in the randomized Tb sublattice. Ideally, such an ordering is expected to give a metamagnetic-like transition in the M-H isotherms at low temperatures. Though such a transition has indeed been observed in Ho(Ni,Fe)$_2$ compounds [15], the M-H isotherms obtained in the Fe-substituted TbNi$_2$ compounds do not show such a transition. Therefore, it seems that, in spite of the similarities in the MCE behavior of Tb(Ni,Fe)$_2$ and Ho(Ni,Fe)$_2$ compounds, there is some difference in their magnetic structures at low temperatures.

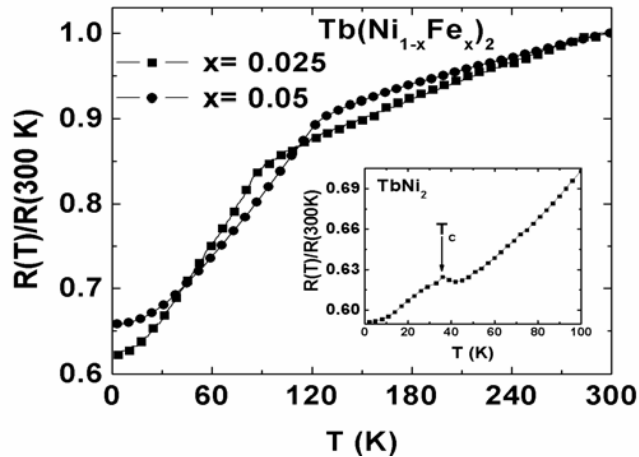

Figure 7. Temperature variation of electrical resistivity, normalized to the value at 300 K, of Tb(Ni,Fe)$_2$ compounds. The inset shows the temperature dependence of normalized resistivity of TbNi$_2$. The arrow in the inset shows the $T_C$ of TbNi$_2$.

Since a change in the magnetic state is expected to reflect in the electrical resistivity behavior, we have also studied the electrical resistivity of these compounds, both as function of temperature as well as the applied magnetic field. The temperature variation of the electrical resistivity of all the compounds, normalized to the value at 300 K, is shown in Figure 7. It can be seen from the figure that all the compounds show metallic character. The onset of magnetic ordering in TbNi$_2$ is characterized by an anomaly in the resistivity whereas in the Fe substituted compounds a change in the slope of the normalized resistivity marks the ordering temperatures. It may be noticed from Figure 7 that the resistivity of TbNi$_2$, near the ordering temperature, initially shows an upturn and then shows a decrease on cooling. Such a behavior is generally expected in antiferromagnetic materials [25]. However, in TbNi$_2$, this behavior of the resistivity may be attributed to the presence of the spin fluctuations. Partial polarization of the Ni -3d band resulting in the formation of non-zero magnetic moment on Ni may be mainly responsible for the enhancement of spin fluctuations. Just below $T_C$, the long range magnetic order sets in which suppresses the fluctuations, thereby causing a drop in the resistivty. It may be recalled here that the temperature variation of heat capacity data of TbNi$_2$ is also indicative of the presence of spin fluctuations just above $T_C$. Such a resistivity behavior has been observed in other intermetallic compounds as well [18].

The field dependence of magnetoresistance (MR), defined as $\left(\dfrac{R(H)-R(0)}{R(0)}\right)$, for all the compounds has been studied in various temperature regions. The MR isotherms of TbNi$_2$ and Tb(Ni$_{0.95}$Fe$_{0.05}$)$_2$, above their ordering temperatures are shown in Figure 8. It can be seen from the figure that near $T_C$, MR value for TbNi$_2$ is negative and its magnitude decreases with increasing temperature and finally becomes positive at higher temperatures. This observation is in contrast with the behavior seen in HoNi$_2$, which showed a positive MR at temperatures close to $T_C$ [15]. The positive MR in HoNi$_2$ was attributed to the dominant contribution from the Lorentz force term [26]. Since the Lorentz force contribution in both these compounds is expected to be almost equal, the presence of negative MR in TbNi$_2$ close to $T_C$ indicates that there is a large negative contribution in this case. The difference between the MR behaviors of HoNi$_2$ and TbNi$_2$ may be due to the difference in the exchange coupling strengths in these two compounds. It is to be noted that the $T_C$ values are 14 K and 36 K for HoNi$_2$ and TbNi$_2$, respectively.

In the Fe substituted compounds, on the other hand, the MR near $T_C$ is strongly negative. While the maximum value of MR for TbNi$_2$ is about 1.6%, it about 6.7% for Tb(Ni$_{0.95}$Fe$_{0.05}$)$_2$, for a field of 50 kOe. The maximum value of MR for Tb(Ni$_{0.975}$Fe$_{0.025}$)$_2$ is found to be ~8.5% for the same field. The presence of large MR in the Fe-substituted compounds may be attributed to suppression of spin fluctuations. It is well known that the MR in many R-TM systems originates due to the suppression of spin fluctuations.

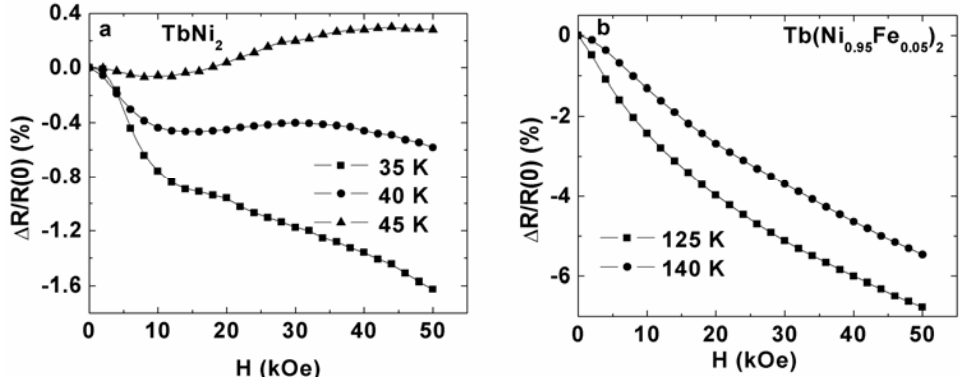

Figure 8 Field dependence of MR in TbNi$_2$ (a) and Tb(Ni$_{0.95}$Fe$_{0.05}$)$_2$ (b) compounds, near their ordering temperatures.

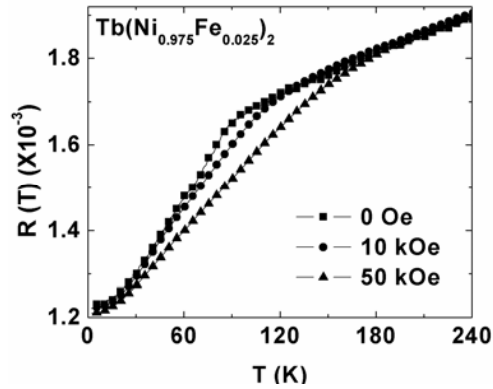

Figure 9 Temperature variation of electrical resistance of Tb(Ni$_{0.975}$Fe$_{0.025}$)$_2$, obtained under various applied magnetic fields.

Figure 9 shows the temperature variation of electrical resistance of Tb(Ni$_{0.975}$Fe$_{0.025}$)$_2$, obtained under various applied magnetic fields. Application of magnetic field reduces the resistance, rendering the MR negative. The suppression of the resistance with the application of the field is maximum near the ordering temperature. It may be noticed that the suppression of the electrical resistance, with the application of the 50 kOe field, starts at about 160 K. This indicates the presence of additional magnetic contribution to electrical resistance, in the Fe-substituted compounds, at temperatures well above $T_C$. We attribute this to the presence of spin fluctuations, as mentioned earlier. A similar behavior has been reported in other intermetallic compounds as well [18]. The fact that the MR at 10 kOe is appreciable only in the vicinity of $T_C$ may be due to the competing effect from the positive Lorentz force contribution and the negative spin fluctuation contribution.

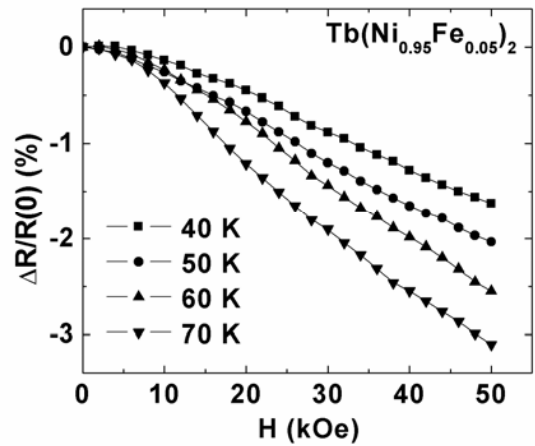

Figure 10 Field dependence of MR in Tb(Ni$_{0.975}$Fe$_{0.025}$)$_2$ at low temperatures.

In order to throw more light on the low temperature behavior of the Fe-substituted compounds, especially near the temperature corresponding to the secondary MCE peak, the field dependence of MR has been estimated and the representative plot of Tb(Ni$_{0.95}$Fe$_{0.05}$)$_2$ is shown in Figure 10. It can be seen from the figure that, in this temperature range, the MR is almost insensitive to the field upto 10 kOe, but increases considerably at higher fields. Since a negative MR could arises due to the suppression of the excess resistivity by the applied field, it is reasonable to assume that some amount of magnetic disorder is present under zero field condition in these compounds, at low temperatures. At this point, it is of importance to recall the signature of randomization seen in the magnetic (dc and ac) and the magnetocaloric results in the case of Fe substituted compounds.

Therefore, we see that the magnetic, magnetocaloric and magnetoresistance behavior in the Fe substituted TbNi$_2$ compounds show considerable similarities with that of Ho(Ni,Fe)$_2$. The anomalous behavior seen in both these systems seems to arise from the unusual magnetic randomness occurring in the rare earth sublattice at low temperatures. Different experimental findings seem to support the proposition of randomization of the R moments. It may be mentioned here that secondary phases and inhomogeneities could also give rise to a similar anomalous behavior. To rule out this possibility, we have carried out microstructural studies, which clearly show that the samples are indeed homogeneous and single phase. In view of this, a detailed neutron diffraction study would be quite essential to probe the exact magnetic structure, especially at low temperatures.

## 4. CONCLUSIONS

In conclusion, the present study shows that MCE and MR show some anomalous behavior in the iron-substituted $TbNi_2$ compounds. It seems that the partial substitution of Fe for Ni leads to the randomization of Tb moments at low temperatures, which gives rise to considerable MCE over a wide range of temperature. We feel that the randomization of Tb sublattice is due to the local anisotropy variations caused by the partial substitution of magnetic Fe in place of nonmagnetic Ni. The present study also shows that the anomalous magnetocaloric and magnetoresistance behavior seen in the present case is similar to those obtained in the Fe substituted $HoNi_2$ compounds.